\documentstyle[prb,aps,epsf]{revtex}
\draft

\begin{document}

\title{Conductance structure in a one-dimensional quantum contact:
dependence on the longitudinal magnetic field}

\author{O.P. Sushkov}

\address{School of Physics, University of New South Wales,\\
 Sydney 2052, Australia}

\maketitle

\begin{abstract}
Very short quantum wires (quantum contacts) exhibit a conductance 
structure at the value of conductance close to $0.7 \times 2e^2/h$. 
The structure was discovered by Thomas {\it et al} (Phys. Rev. Lett. 
{\bf 77}, 135 (1996)). Dependence of the structure on the longitudinal 
magnetic field was studied in the same experiment. This dependence
has clearly demonstrated that the effect is related 
to the electron spin. In the recent work (Sushkov, Phys. Rev. B {\bf 64}, 
(2001)) it was suggested 
that the effect is caused by the development of the charge density
wave within the contact.  The  many-body Hartree-Fock calculation 
has confirmed the picture. The exchange electron-electron interaction
is crucial for this calculation, so in agreement with experiment the effect 
is intrinsically related to the spin. However the dependence on the 
magnetic field has not been calculated yet. In the present paper  
the many-body Hartree-Fock calculation of conductance in the presence of the 
longitudinal magnetic field is performed.
\end{abstract}


The quantized conductance $G=nG_2$, $n=1,2,3,...$, $G_2=2e^2/h$,
through a narrow quantum point contact was
discovered in 1988 \cite{Wharam,Wees}. This quantization can be  
understood within a one-dimensional (1D)
non-interacting electron gas picture, see e.g. Ref.\cite{But}.
In the present work we are interested in a deviation from the
integer quantization. This deviation, the  so called 
``0.7 structure'' has been found in experimental works \cite{Thomas,Thomas1}.
The structure is a shoulder-like feature or a narrow plateau at
$G\approx 0.7G_2$. More recent work demonstrates that there are some
above barrier excitations related to the structure \cite{Kristensen}, and
that the structure evolves down to $G\approx 0.5G_2$ in longer quantum 
contacts \cite{Reilly}.

Dependence of the structure on the longitudinal magnetic field
was studied already in the pioneering work \cite{Thomas}. This study 
clearly demonstrated that the effect is somehow related to the electron 
spin. There have been numerous attempts to explain the ``0.7 structure'' 
by spontaneous magnetization of the 1D quantum wire
\cite{Chuan,Calmels,Zabala,V'yurkov,Spivak,Bychkov}, or  by the formation of a 
two-electron bound state with nonzero total spin \cite{Flamb,Rejec}.
However these scenarios contradict to the rigorous Lieb-Mattis theorem 
\cite{Lieb} that claims that a 1D many-body system with a potential 
interaction has zero spin in the ground state. 

In the recent work \cite{Sushkov} I have suggested that the effect is caused 
by the development of the charge density wave within the 
contact.  This wave is a precursor for 1D Wigner
crystallization \cite{Glaz}. The total spin of the system without a
magnetic field is zero in agreement with the Lieb-Mattis theorem \cite{Lieb}. 
The many-body Hartree-Fock calculation \cite{Sushkov} has confirmed the 
picture and reproduced the ``0.7 structure''. The exchange electron-electron 
interaction is crucial for this 
calculation, in this sense the effect is intrinsically related to spin. 
However to probe explicitly the spin dependence one has to follow the path 
of the experiment \cite{Thomas,Thomas1} and to calculate the
quantum contact conductance in the presence of the longitudinal magnetic
field. Such calculation is performed in the present work.

 The present work employs the same technique that has been developed in
Ref. \cite{Sushkov}. This is the Hartree-Fock (HF) method combined with the
fictitious gauge field method. However there are some complications that
are due to the magnetic field. Let us first formulate the idea of the 
approach \cite{Sushkov}.
In independent particle approximation, i.e. in the case of an ideal electron
gas, a calculation of the conductance for a given transverse channel is 
straightforward
\begin{equation}
\label{GT}
G={{2e^2}\over{h}} T,
\end{equation}
where $T$ is the barrier transmission probability at Fermi energy 
\cite{Land,But}.
In case of interacting particles this formula is also valid because
before and after the potential barrier the density of electrons
is high enough, and hence the interaction is negligible.
However one cannot use the single particle description to 
calculate the transmission probability $T$ because in the vicinity of the 
barrier the electron density is low, and hence the many-body effects are very 
important.
To calculate the transmission probability $T$ the following method is
applied. Consider the liquid of electrons on a 1D ring with a potential 
barrier somewhere on the ring.
There is no  current in the ground state of the system. Now let us apply a
magnetic flux through the ring. This flux induces the electric current.
Note that this is not a real magnetic field, this is a fictitious gauge
field that generates the current without applying any voltage.
On the one hand the current is related to the barrier transmission 
probability $T$. On the other hand, the current can be calculated using 
the HF method. This allows to find $T$ with account of many-body effects.

Now we have to repeat the same calculation with addition of the real
longitudinal magnetic field $B$. This field does not influence an
orbital dynamics, but it splits Fermi levels for electrons with spin up
and spin down. So we have to perform the HF calculation in the sector
with nonzero total spin. According to the experimental data 
\cite{Thomas,Thomas1} the dependence of the conductance  on the gate
voltage evolves with the magnetic field. The field influences the 0.7
structure, and at the value of the field $B\approx 5-10 \ T$ the conductance
evolves to the double step function that one should expect for a simple
spin split band. The effective value of the electron $g$ factor depends
on the width of the channel.
For a wide channel it is close to the bulk limit $g \approx 0.4$, and
for a narrow channel it is $g \approx 1$, see Refs.\cite{Thomas,Thomas1}. 
There is only one channel in the quantum contact we are 
interested in, so the contact is narrow.
On the other side the electrons incident on the contact are coming from
the bulk and the degree of their polarization is determined by the bulk.
Therefore for the estimate we take the intermediate value $g \sim 0.7$. Then 
the energy splitting corresponding to $B=5T$ is
\begin{equation}
\label{split}
\Delta E = g\mu_B B=2 \ 10^{-4}eV=2 \ 10^{-2} E_{unit}.
\end{equation}
We use atomic units, so distances are measured in unites of Bohr
radius, $a_B=\epsilon\hbar^2/m e^2$, and energies are measured in units
of $E_{unit}=me^4/\hbar^2\epsilon^2$, where $m$ is the effective electron
mass and $\epsilon$ is the dielectric constant. 
For experimental conditions of works \cite{Thomas,Thomas1,Kristensen,Reilly} 
these values are the following: $a_B \approx 10^{-2}\mu m$, 
$E_{unit}\approx 10^{-2}eV$.
The energy scale (\ref{split}) corresponds to the temperature 
$T \approx 2 \ K$. It is interesting to note that the ``0.7 structure'' 
depends on the temperature exactly on this scale 
\cite{Thomas,Thomas1,Kristensen,Reilly}. At this stage this is just 
an observation. In the present  work we consider only the case of
zero temperature.

The Hamiltonian of the many body system that we consider is of the form
\begin{equation}
\label{Hmb}
H=\sum_i\left[{{(p_i-{\cal A})^2}\over{2}}+U(x_i)-2g\mu_B{\bf B}\cdot{\bf s_i}
\right]+{1\over{2}}\sum_{i,j}V(x_i,x_j),
\end{equation}
where indexes $i$ and $j$ numerates electrons, ${\bf s_i}$ is the electron 
spin, $x_i$ is the periodic coordinate on the ring of length $L$ ($0<x<L$), 
${\cal A}=\pi/2L$ is the fictitious gauge field, and $B$ is the longitudinal 
magnetic field. 
The electron-electron Coulomb repulsion is of the form
\begin{equation}
\label{Vij}
V(x,y)={1\over{\sqrt{a_t^2+D^2(x,y)}}},
\end{equation}
where $a_t\approx 2$ is the effective width of the transverse channel, and 
$D(x,y)$ is the length of the shortest arc  between the points $x$ and $y$
on the ring, for more details see Ref. \cite{Sushkov}. 
To model the gate potential \cite{Thomas,Thomas1,Kristensen,Reilly}
we use the following formula for the potential barrier 
\begin{equation}
\label{UU}
U(x)={{U_0}\over{e^{(|x|-l/2)/d}+1}}.
\end{equation}
In the present work we take $l=10$ and $d=1$. The plot of $U(x)$ is shown 
in Fig.1 by a solid line
\begin{figure}[h]
\vspace{-10pt}
\hspace{-35pt}
\epsfxsize=10.cm
\centering\leavevmode\epsfbox{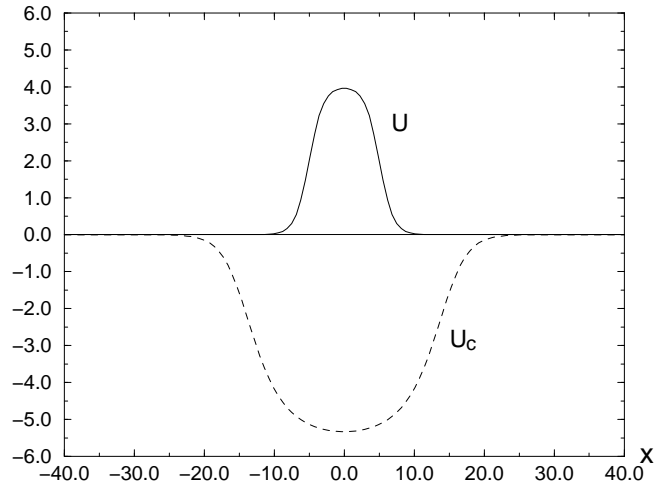}
\vspace{-10pt}
\caption{\it {The solid line shows the external potential (\ref{UU}) at 
$U_0=4$. The dashed line shows the compensating potential $U_c(x)$.
}}
\label{Fig1}
\end{figure}
\noindent
It was demonstrated in Ref.\cite{Sushkov} that to find the barrier transmission
probability at Fermi energy with account of many-body effects one has to 
solve the many-body problem (\ref{Hmb}) twice: without the barrier and with 
the barrier. The ratio of electric currents squared gives the transmission
probability $T=(J_U/J_0)^2$. This formula is valid without an external magnetic
field. Repeating considerations of Ref.\cite{Sushkov} one can prove that
with the magnetic field, i.e. with the spin splitting, the probability is
given by
\begin{equation}
\label{TT}
T={1\over{2}}\left({{J_{U\uparrow}}\over{J_{0\uparrow}}}\right)^2+
{1\over{2}}\left({{J_{U\downarrow}}\over{J_{0\downarrow}}}\right)^2,
\end{equation}
where $J_{\uparrow \downarrow}$ is the electric current of electrons
with spin up and spin down correspondingly.
This formula gives the probability and hence due to eq. (\ref{GT}) it
solves the many-body problem of conductance.

To solve the many-body problem (\ref{Hmb}) we use the Hartree-Fock (HF)
approximation. In the HF approximation the many body wave function is 
represented in the form of the Slater determinant of single particle orbitals  
$\varphi_{i\sigma}(x)$. 
The index $i$ shows the coordinate state of the orbital, and the index
$\sigma = \pm 1/2$ shows the spin state of the orbital.
Each orbital obeys the equation
\begin{equation}
\label{hf}
\hat{h}\varphi_{i\sigma}=\epsilon_{i\sigma}\varphi_{i\sigma},
\end{equation}
where $\epsilon_{i\sigma}$ is the single particle energy and
$\hat{h}$ is the HF Hamiltonian
\begin{eqnarray}
\label{Hhf}
\hat{h}\varphi_{i\sigma}(x)&=&\left({{(p-{\cal A})^2}\over{2}}+U_{eff}(x)
-g\mu_BB\sigma\right)\varphi_{i\sigma}(x)
-\sum_j\int\varphi_{j\sigma}^*(y)\varphi_{i\sigma}(y)V(x,y)dy\varphi_{j\sigma}
(x),\\
U_{eff}&=&U(x)+ \sum_{j\sigma}\int|\varphi_{j\sigma}(y)|^2 V(x,y)dy
+U_c(x).\nonumber
\end{eqnarray}
The summations are performed over all filled orbitals. The compensating 
potential at this stage is zero, $U_c=0$. We will use the compensating
potential later to deal with some computational problems.

For computations we use a finite grid.  
In the grid modification of the Hamiltonian (\ref{Hhf}) the kinetic 
energy $(p-{\cal A})^2\varphi$ is replaced by
$\left[2|\varphi(n)|^2-\varphi^*(n+1)e^{iAh}\varphi(n)
-\varphi^*(n)e^{-iAh}\varphi(n+1)\right]/2h^2$. Here $h$ is the step of the
grid and $\varphi(n)$ is the wave function on the site $n$ of the grid.
The electric current corresponding to the grid Hamiltonian is
\begin{equation}
\label{jg}
J_{\sigma}=-\sum_j{{i}\over{2h}}\left[\varphi_{j\sigma}^*(n)e^{iAh}
\varphi_{j\sigma}(n+1)-
\varphi_{j\sigma}^*(n+1)e^{-iAh}\varphi_{j\sigma}(n)\right].
\end{equation}
The current is conserved because of the gauge invariance of HF equations.

In the computations we use the grid of 400 points and the total number of 
electrons  $N=N_{\uparrow}+N_{\downarrow}=158$. Because of the computational
time limitation it is very hard to increase substantially these numbers.
The transmission probability versus the external potential $U_0$, see
eq. (\ref{UU}), is shown in Fig.2. It has been calculated at zero magnetic 
field, i. e. at $N_{\uparrow}=N_{\downarrow}=79$,  for the ring of length 
$L=80$.
\begin{figure}[h]
\vspace{-10pt}
\hspace{-35pt}
\epsfxsize=10.cm
\centering\leavevmode\epsfbox{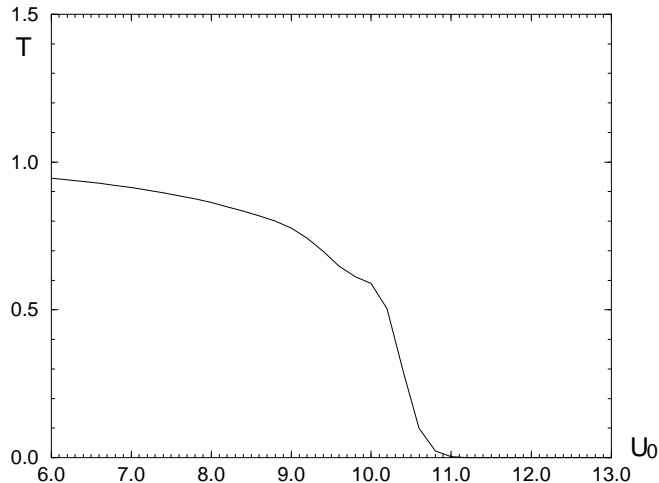}
\vspace{-10pt}
\caption{\it {The transmission probability versus the external potential
barrier height $U_0$. Length of the ring is $L=80$.
}}
\label{Fig2}
\end{figure}
\noindent
The Fig.2 clearly indicates the ``0.7 structure''. It was discussed
in detail in Ref. \cite{Sushkov} that the structure is due to the Coulomb
exchange interaction on the barrier. 

To find the effect of the longitudinal magnetic field one has to repeat the 
above calculation with the magnetic field. In general terms it is 
straightforward, but there is a serious technical complication.
In the above calculation the density of electrons outside the
barrier is $n\approx 158/80 \approx 2$. The corresponding
Fermi momentum is $p_F=\sqrt{2\pi n} \approx 3.5$.
The electron spectrum is discreet with 
energy splitting $\delta \epsilon = p_F \pi/L \approx 0.1$. This is ten times
larger than the splitting (\ref{split}) due to the magnetic field
5T. Thus the system is not sensitive to the magnetic field  up to $B\sim 50T$.
Certainly it is not a physical effect, it is a byproduct of the finite
ring method used in the calculation. To resolve the problem in a 
straightforward way one has to
increase the length of the ring by at least an order of magnitude. The density
of electrons must be substantially higher than the critical density 
\cite{Glaz}. It means that the total number of electrons must be at least
1000-2000. This is computationally impossible.

To resolve the problem we use the following combination of the means:
1)Increase the length up to $L=260$.
2)Make the electron-electron interaction (\ref{Vij}) dependent on the
position on the ring: 
\begin{equation}
\label{VVV}
V(x,y)\to V_d(x,y)={{
V(x,y)}\over{\left(e^{(|x|-15)/2}+1\right)
\left(e^{(|y|-15)/2}+1\right)}},
\end{equation}
3) Introduce into eq. (\ref{Hhf}) an additional compensating 
 potential $U_c(x)$.
The interaction $V_d$ coincides with the Coulomb interaction $V(x,y)$ in
the interval [-15,15], and $V_d$ vanishes outside of this interval.
Since there is no interaction at large distances we
can have a relatively low density of electrons at $|x|>15$.
There is a prize for the translationally noninvariant interaction $V_d$.
Unfortunately the interaction produces the unphysical Coulomb barrier. To 
compensate the 
unphysical barrier we introduce the compensating potential $U_c(x)$ adjusted 
in such a way that in absence of the real potential barrier ($U_0=0$, see eq.
(\ref{UU})) the density of electrons at the contact, $-15 < x < 15$, is
$n\approx 1$. The compensating potential $U_c(x)$ is shown in Fig.1 by the
dashed line. The density of electrons $n(x)$ obtained by solution
of HF eqs. with $V_d$ and $U_c$ but without any external barrier ($U_0=0$)
 is shown in Fig.3.
\begin{figure}[h]
\vspace{-10pt}
\hspace{-35pt}
\epsfxsize=10.cm
\centering\leavevmode\epsfbox{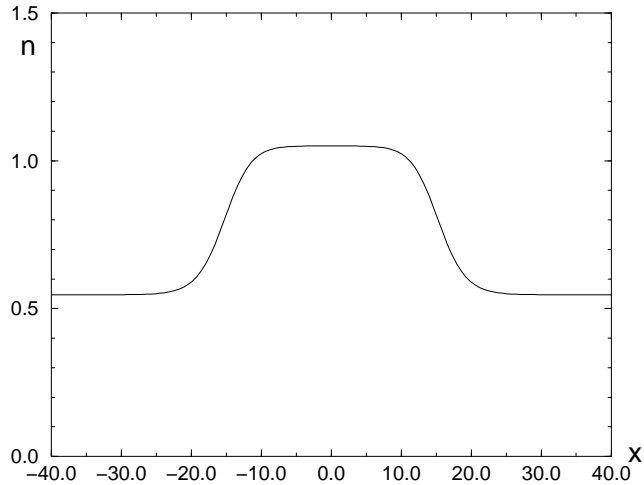}
\vspace{-10pt}
\caption{\it {Selfconsistent density of electrons calculated with the
modified Coulomb interaction $V_d$ and with the compensating potential $U_c$.
The external potential is zero, $U(x)=0$. The length of the ring
is $L=260$.
}}
\label{Fig3}
\end{figure}
\noindent
The energy splitting between single particle levels in this case is 
$\delta\epsilon \approx 0.01$. The spin flip of a single electron
creates an additional spin up electron and an  additional spin down hole, so it
``costs'' $2\epsilon\approx 0.02$. This must be smaller than the
magnetic splitting (\ref{split}).
So the minimal magnetic field we can consider is 5T.

Hartree-Fock calculations for the long ring (L=260) with
the modified Coulomb interaction (\ref{VVV}) and with the corresponding
compensating potential $U_c$ are absolutely similar to that described
before. In this case the effective potential $U_{eff}$, see eq. 
(\ref{Hhf}), is not constant even without any external barrier (\ref{UU}).
However this effective potential is smooth and well below the Fermi
energy. Therefore it does not influence the transmission probability.
So in equation (\ref{TT}) for the transmission probability 
the current $J_0$ corresponds to the zero external potential, and the
current $J_U$ corresponds to some given value of the external potential
$U_0$.
The transmission probability at zero magnetic field 
(i.e. $N_{\uparrow}=N_{\downarrow}=79$) versus the barrier height $U_0$
is plotted in Fig.4 by the solid line.
\begin{figure}[h]
\vspace{-10pt}
\hspace{-35pt}
\epsfxsize=10.cm
\centering\leavevmode\epsfbox{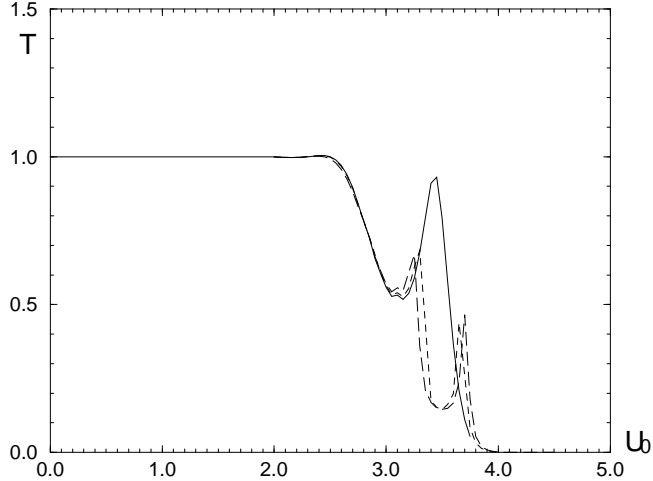}
\vspace{-10pt}
\caption{\it {The transmission probability versus the external potential
barrier height $U_0$. This is the Hartree-Fock calculation on the ring of
length $L=260$ with the modified Coulomb interaction. The solid line 
corresponds to zero magnetic
field. The long-dashed line corresponds to the field $B=5T$,
and the dashed line corresponds to $B=10T$.
}}
\label{Fig4}
\end{figure}
\noindent
One can see that after smearing the average position of the structure 
is close to $T\approx 0.7$. However compared to Fig.2 the structure is 
much more  pronounced.
This happens because the modified Coulomb interaction (\ref{VVV})
underestimates the Hartree screening, at the same 
time the exchange interaction responsible for the structure is not influenced
by the modification.
 Underestimation of the Hartree screening leads also
to the shift in the position of the step: it is at $U_0 \sim 3.5$ in
Fig.4 and at $U_0 \sim 10.5$ in Fig.2 (full screening).
However all these details are not that important, the ``0.7 structure''
qualitatively and to large extent quantitatively  is described.
What is important is that the modification gives us
the possibility to work with the magnetic field.
The result of calculation for $N_{\uparrow}=80$ and $N_{\downarrow}=78$
is shown in Fig. 4 by the long dashed line. As it was discussed before
it corresponds to the field $B=5T$.
The result for $N_{\uparrow}=81$ and $N_{\downarrow}=77$
is shown by the dashed line. It corresponds to the field $B=10T$.
In agreement with the data \cite{Thomas,Thomas1} the field $B\sim 5T$
shifts the position of the structure close to $T\approx 0.5$.
Further increasing of the field practically does not influence the
picture.

It is very illustrative to perform also  calculations in the Hartree
approximation, i. e. calculations without exchange interaction. The results 
for $B=0$ (solid line), $B=5T$ 
(long-dashed line), $B=10T$ (dashed line) are plotted in Fig. 5.
\begin{figure}[h]
\vspace{-10pt}
\hspace{-35pt}
\epsfxsize=10.cm
\centering\leavevmode\epsfbox{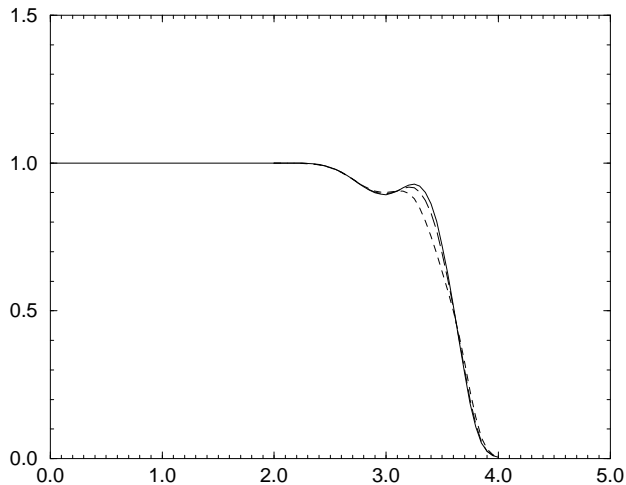}
\vspace{-10pt}
\caption{\it {The transmission probability versus the external potential
height $U_0$  calculated in the Hartree approximation.
 The solid line corresponds  to zero magnetic
field. The long-dashed line corresponds to the field $B=5T$,
and the dashed line corresponds to $B=10T$.
}}
\label{Fig5}
\end{figure}
\noindent
As one should expect neither ``0.7 structure'' no magnetic field
dependence appear in the Hartree calculation.

In conclusion we have performed the many-body Hartree-Fock calculation
of conductance in the presence of the longitudinal magnetic field.
Results of the calculation are shown in Fig.4 where the transmission
probability (the conductance in units $2e^2/h$) is plotted versus the
potential barrier height that models the gate potential.
The solid line corresponds to zero magnetic
field, the long-dashed line corresponds to the field $B=5T$,
and the dashed line corresponds to $B=10T$.
There are some computational limitations related to work with the magnetic 
field. To overcome these limitations we had to modify
the electron-electron Coulomb interaction at large distances. This is why the 
structure in
conductance in Fig.4 is more pronounced than the structure in the calculation
without the modification shown in Fig.2.
Our calculation demonstrates that the magnetic field $B\sim 5T$ shifts 
the position 
of the structure close to $T\approx 0.5$. Further increasing of the field 
practically does not influence the picture.
These results are in qualitative and to a large extent quantitative
agreement with the experimental data \cite{Thomas,Thomas1}.
The agreement gives further confirmation of the explanation of the
``0.7 structure'' suggested in Ref.\cite{Sushkov}.

\end{document}